\documentclass[authoryear,review,12pt]{elsarticle}

 \usepackage{graphicx,color}
\usepackage{natbib}
\usepackage{appendix}

\journal{Icarus}

\usepackage{amsmath,amssymb}

\newcommand\unit[1]{\ \mathrm{#1}}

\def\cm{\unit{cm}}
\def\m{\unit{m}}
\def\km{\unit{km}}
\def\AU{\unit{AU}}

\def\gm{\unit{g}}
\def\My{\unit{My}}
\def\Gy{\unit{Gy}}

\def\s{\unit{s}}
\def\yr{\unit{y}}

\begin{document}

\begin{frontmatter}



\title{Dynamical erosion of the asteroid belt and implications for large impacts in the inner solar system}


\author{David A. Minton and Renu Malhotra} 

\address{Lunar and Planetary Laboratory, The University of Arizona, Tucson, AZ 85721}

\begin{abstract}
The cumulative effects of weak resonant and secular perturbations by the major planets produce chaotic behavior of asteroids on long timescales.
Dynamical chaos is the dominant loss mechanism for asteroids with diameters $D\gtrsim10\km$ in the current asteroid belt.
In a numerical analysis of the long term evolution of test particles in the main asteroid belt region,
we find that the dynamical loss history of test particles from this region is well described with a logarithmic decay law.
In our simulations the loss rate function that is established at $t\approx1\My$ persists with little deviation to at least $t = 4\Gy$.
Our study indicates that the asteroid belt region has experienced a significant amount of depletion due to this dynamical erosion --- having lost as much as $\sim50\%$ of the large asteroids --- since $1\My$ after the establishment of the current dynamical structure of the asteroid belt.  
Because the dynamical depletion of asteroids from the main belt is approximately logarithmic, an equal amount of depletion occurred in the time interval $10$--$200\My$ as in $0.2$--$4\Gy$, roughly $\sim30\%$ of the current number of large asteroids in the main belt over each interval. 
We find that asteroids escaping from the main belt due to dynamical chaos have an Earth impact probability of $\sim0.3\%$.
Our model suggests that the rate of impacts from large asteroids has declined by a factor of $3$ over the last $3\Gy$, and that the present-day impact flux of $D>10\km$ objects on the terrestrial planets is roughly an order of magnitude less than estimates currently in use in crater chronologies and impact hazard risk assessments.

\end{abstract}


\end{frontmatter}



\section{Introduction}
\label{sec:asteroid_loss-intro}
The main asteroid belt spans the $\sim2$--$4\AU$ heliocentric distance zone that is sparsely populated with rocky planetesimal debris.
Strong mean motion resonances with Jupiter in several locations in the main belt cause asteroids to follow chaotic orbits and be removed from the main belt~\citep{Wisdom:1987p206}.
These regions are therefore emptied of asteroids over the age of the solar system, forming the well-known Kirkwood gaps~\citep{Kirkwood:1867p48}.
In addition to the well known low-order mean motion resonances with Jupiter that form the Kirkwood gaps, there are numerous weak resonances that cause long term orbital chaos and transport asteroids out of the main belt~\citep{Morbidelli:1999p127}.
A very powerful secular resonance that occurs where the pericenter precession rate of an asteroid is nearly the same as that of one of the solar system's eigenfrequencies, the $\nu_6$ secular resonance, lies at the inner edge of the main belt~\citep{Williams:1981p532}.

The many resonances found throughout the main asteroid belt are largely responsible for maintaining the Near Earth Asteroid (NEA) population.  
Non-gravitational forces, such as the Yarkovsky effect, cause asteroids to drift in semimajor axis into chaotic resonances whence they can be lost from the main belt~\citep{Opik:1951p1505,Vokrouhlicky:2000p706,Farinella:1998p18,Bottke:2000p82}.
The Yarkovsky effect is size-dependant, and therefore smaller asteroids are more mobile and are lost from the main belt more readily than larger ones. 
Asteroids with $D\lesssim 10\km$ have also undergone appreciable collisional evolution over the age of the solar system~\citep{Geissler:1996p83,Cheng:2004p65,Bottke:2005p107}, and collisional events can also inject fragments into chaotic resonances~\citep{Gladman:1997p88}.
These processes (collisional fragmentation and semimajor axis drift followed by injection into resonances) have contributed to a quasi steady-state flux of small asteroids ($D\lesssim10\km$) into the terrestrial planet region and are responsible for delivering the majority of terrestrial planet impactors over the last $\sim3.5\Gy$~\citep{Bottke:2000p82,Bottke:2002p113,Bottke:2002p117,Strom:2005p80}.

In contrast, most members of the population of $D\gtrsim30\km$ asteroids have existed relatively unchanged, both physically and in orbital properties, since the time when the current dynamical architecture of the main asteroid belt was established:  the Yarkovsky drift is negligble and the mean collisional breakup time is $>4\Gy$ for $D\gtrsim30\km$ asteroids; asteroids with diameters between $\sim10$--$30\km$ have been moderately altered by collisional and non-gravitational effects. 
However, as we show in the present work, the population of large asteroids ($D\gtrsim10$--$30\km$) is also subject to weak chaotic evolution and escape from the main belt on gigayear timescales.  
By means of numerical simulations, we computed the loss history of large asteroids in the main belt. 
We also computed the cumulative impacts of large asteroids on the terrestrial planets over the last $\sim3\Gy$.

Our present study is additionally motivated by the need to understand better the origin of the present dynamical structure of the main asteroid belt.
The orbital distribution of large asteroids that exist today was determined by dynamical processes in the early solar system.
\cite{Minton:2009p280} showed that large asteroids do not uniformly fill regions of the main belt that are stable over the age of the solar system.  
The Jupiter-facing boundaries of some Kirkwood gaps are more depleted than the Sunward boundaries, and the inner asteroid belt is also more depleted than a model asteroid belt in which only gravitational perturbations arising from the planets in their current orbits have sculpted an initially uniform distribution of asteroids.
\cite{Minton:2009p280} showed that the pattern of depletion observed in the main asteroid belt is consistent with the effects of resonance sweeping due to giant planet migration that is thought to have occurred early in solar system history~\citep{Fernandez:1984p61,Malhotra:1993p244,Malhotra:1995p79,Hahn:1999p122,Gomes:2005p51}, and that this event was the last major dynamical depletion event experienced by the main belt. 
The last major dynamical depletion event in the main asteroid belt likely coincided with the so-called Late Heavy Bombardment (LHB) $\sim3.9\Gy$ ago as indicated by the crater record of the inner planets and the Moon~\citep{Strom:2005p80}.

Knowledge about the distribution of the asteroids in the main belt just after that depletion event may help constrain models of that event.
Quantifying the dynamical loss rates from the asteroid belt also help us understand the history of large impacts on the terrestrial planets.
Motivated by these considerations, in this paper we explore the dynamical erosion of the main asteroid belt, which is the dominant mechanism by which large asteroids have been lost during the post-LHB history of the solar system. 

We have performed n-body simulations of large numbers of test particles in the main belt region for long periods of time ($4\Gy$ and $1.1\Gy$).
We have derived an empirical functional form for the population decay and for the dynamical loss rate of large main belt asteroids.  
Finally, we discuss the implications of our results for the history of large asteroidal impacts on the terrestrial planets.

\section{Numerical simulations}
\label{sec:asteroid_loss-simulations}
Our long term orbit integrations of the solar system used a parallelized implementation of a second-order mixed variable symplectic mapping known as the Wisdom-Holman Method~\citep{Wisdom:1991p139,Saha:1992p42}, where only the massless test particles are parallelized and the massive planets are integrated in every computing node.
Our model included the Sun and the planets Mars, Jupiter, Saturn, Uranus, and Neptune.
All masses and initial conditions were taken from the JPL Horizons service\footnote{see http://ssd.jpl.nasa.gov/?horizons} on July 21, 2008.
The masses of Mercury, Venus, Earth, and the Moon were added to the mass of the Sun.

\begin{figure}
\center
\resizebox{\textwidth}{!}{\includegraphics{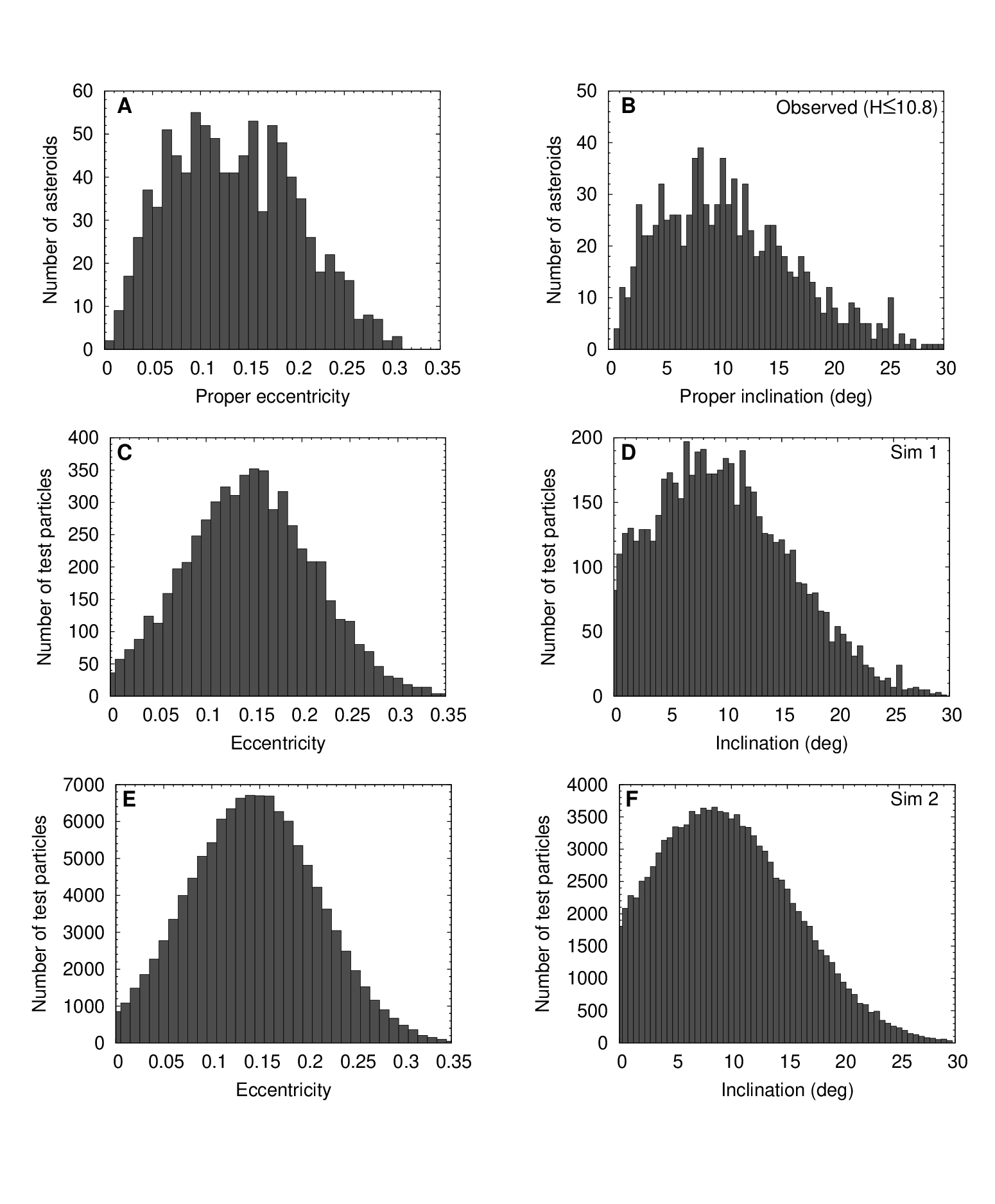}}
\caption{Eccentricity and inclination distributions.
\textbf{a)},\textbf{b)} The distributions of the 931 observed bright ($H\leq10.8$) asteroids in the main belt that are not members of collisional families, from the AstDys online data service~\citep{Knezevic:2003p74, Nesvorny:2006p2242}.
\textbf{c)},\textbf{d)} The initial $e$ and $i$ distributions for Sim~1 (the 5760 particle simulation).
\textbf{e)},\textbf{f)} The initial $e$ and $i$ distributions for Sim~2 (the 115200 particle simulation).}
\label{f:aedist}      
\end{figure}

Test particle asteroids were given eccentricity and inclination distributions similar to the observed main belt, but a uniform distribution in semimajor axis.
The initial eccentricity distribution of the test particles was modeled as a Gaussian with the peak at $\mu=0.15$ and a standard deviation of 0.07, a lower cutoff at zero, and an upper cutoff above any value which would lead to either a Mars or Jupiter-crossing orbit, whichever was smaller.
The initial inclination distribution was modeled as a Gaussian with the peak $\mu=8.5^\circ$ and standard deviation of $7^\circ$, and a lower cutoff at $0^\circ$.
The other initial orbital elements (longitude of ascending node, longitude of perihelion, and mean anomaly) were uniformly distributed.
The eccentricity and inclination distributions of the adopted initial conditions and those of the observed asteroids of absolute magnitude $H\leq10.8$ are shown in Fig.~\ref{f:aedist}.

Mars was the only terrestrial planet integrated in our simulations. 
Despite its small mass, Mars has a significant effect on the dynamics of the inner asteroid belt due to numerous weak resonances, including three-body Jupiter-Mars-asteroid resonances~\citep{Morbidelli:1999p127}.
Two simulations were performed: Sim~1, with 5760 test particles integrated for $4\Gy$, and Sim~2, with 115200 test particles integrated for $1.1\Gy$. 
In each of these simulations an integration step size of $0.1\yr$ was used.  
Particles were considered lost if they approached within a Hill radius of a planet, or if they crossed either an inner boundary at $1\AU$ or an outer boundary at $100\AU$.

We define time $t=0$ as the epoch when the current dynamical architecture of the main asteroid belt and the major planets was established. 
What we mean by this is the time at which any primordial mass depletion and excitation has already taken place~\citep[see][]{OBrien:2007p95}, and any early orbital migration of giant planets has finished~\citep{Fernandez:1984p61,Malhotra:1993p244,Strom:2005p80,Gomes:2005p51,Minton:2009p280}.
At this epoch the main belt would have already had its eccentricity and inclination distributions excited by some primordial process, and its semimajor axis distribution shaped by early planet migration. 
Therefore, the $e$ and $i$ distributions at $t=0$ likely resembled those of the present-day asteroid belt, although subsequent long-term evolution likely altered them somewhat from their primordial state.
Current understanding of planetary system formation suggests that the epoch prior to when we define $t=0$ could have been several million to several hundred million years subsequent to the formation of the first solids in the protoplanetary disk; the first solids have radiometrically determined ages of $4.567\Gy$~\citep{Russell:2006p1433}.

\section{Main asteroid belt population evolution}
\label{sec:asteroid_loss-results}
\begin{figure}
\center
\resizebox{\textwidth}{!}{\includegraphics{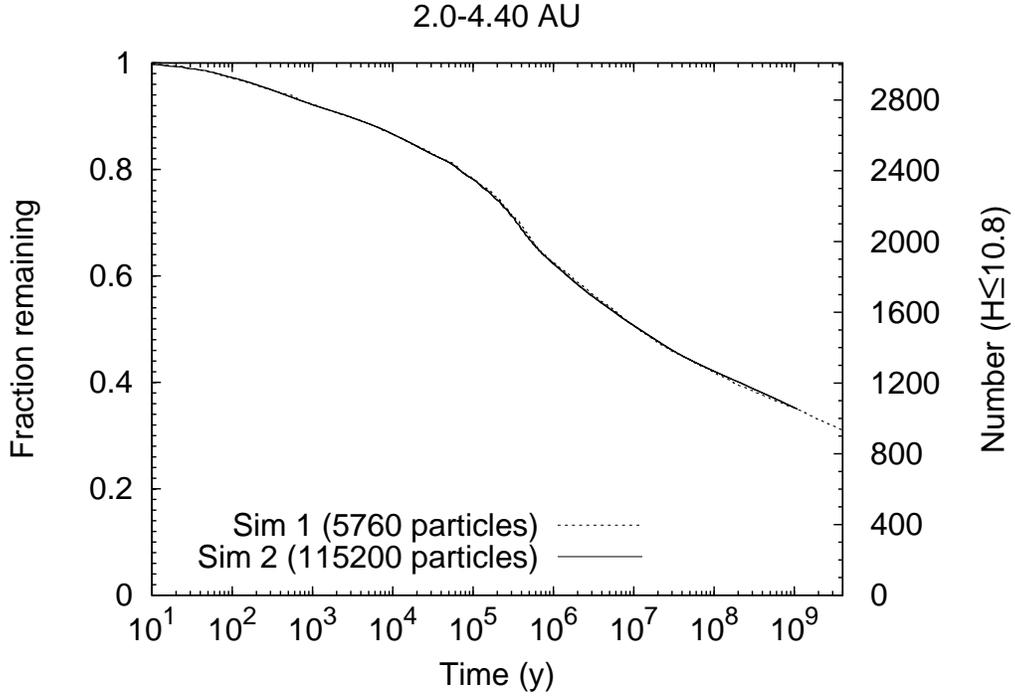}}
\caption{Loss history of test particles in the main asteroid belt region of the solar system from both Sim~1 (5760 particles for $4\Gy$) and Sim~2 (115200 particles for $1.1\Gy$).
The left-hand axis is the fraction of the original test particle population that have survived the simulation at a given time.
The right-hand axis is the estimated number of large asteroids in the asteroid belt, and is computed by equating the fraction remaining at $t=4\Gy$ with the number of observed $H\leq10.8$ asteroids.
The observational sample used is the 931 asteroids with $H\leq10.8$ excluding members of collisional families.}
\label{f:lhistory}      
\end{figure}

The loss history of particles from Sim~1 and Sim~2 are shown in Fig.~\ref{f:lhistory}.  
The loss histories are nearly indistinguishable over the $1.1\Gy$ length of Sim~2.  
The loss history appears to go through two phases.  
The first phase, lasting until $\sim1\My$, is characterized by a rapid loss of particles from highly unstable regions, such as the major Kirkwood gaps and the $\nu_6$ secular resonance.
The slope of the loss rate on a log-linear scale changes rapidly between $0.3$--$1\My$ until the second phase is reached, which lasts from $1\My$ until at least the end of Sim~1 at $4\Gy$.
The slope of the loss rate on a log-linear scale continues to change during the second phase, but only over much longer timescales and by a much smaller amount than during the first phase. 

\begin{figure}
\center
\resizebox{\textwidth}{!}{\includegraphics{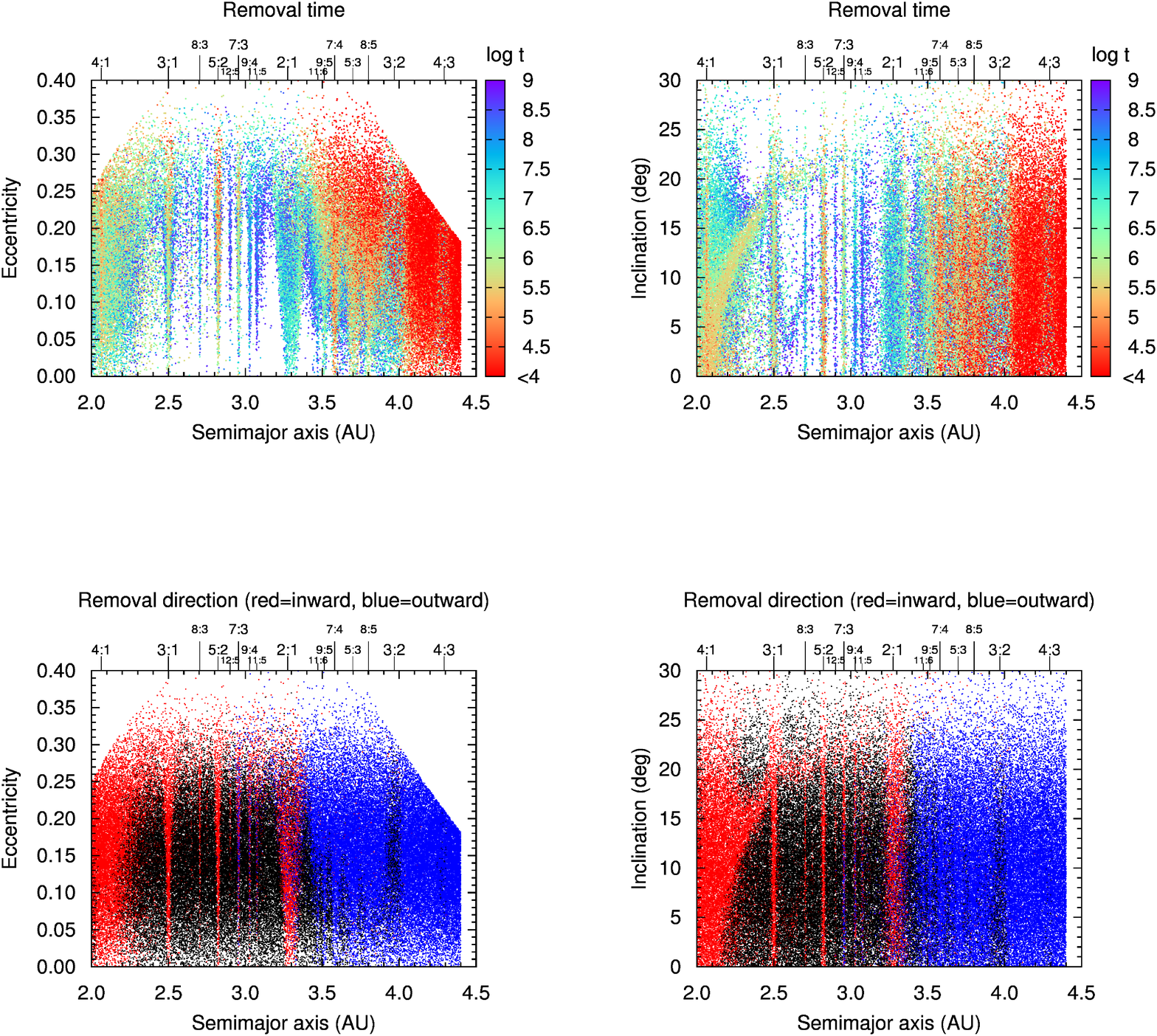}}
\caption{Removal statistics of the test particles in Sim~2 (115200 particles) as function of their initial orbital elements.  
In the upper panels, the points are colored to indicate their lifetime in the simulation, with red being the shortest-lived particles and blue being the longest-lived particles (particles surviving at the end of the simulation were removed for clarity).
In the lower panels, the points are colored to indicate the direction in which they are lost:  red indicates loss due to either a close encounter with Mars or removal at the inner boundary at $1\AU$, blue indicates loss due either a close encounter with a giant planet or removal at the outer boundary at $100\AU$, and black indicates particles that survived the entire $1.1\Gy$ simulation.}
\label{f:cdist}      
\end{figure}

The particle removal times and particle fates (whether they become inward-going Mars-crossers, or outward-going Jupiter-crossers) for Sim~2 are both shown in Fig.~\ref{f:cdist}.
We find that the particles that are lost during the initial $1\My$ (the red to light-green points) are generally those with high initial eccentricity, particles from the $\nu_6$ resonance (appearing as a curving yellow band in the semimajor axis vs. inclination plot), and particles in the strongly chaotic mean motion resonances with Jupiter (the Kirkwood gaps).
These maps are similar to those produced by \cite{Michtchenko:2009p2516}, however their maps were coded by spectral number (more chaotic orbits having a larger spectral number) using $4.2\My$ integrations.
The apparent rapid change in slope around $10^5$--$10^6$ years is likely due mostly to the emptying of asteroids from the $\nu_6$ resonance region (see also the upper right-hand panel of Fig.~\ref{f:cdist}).
Also, in the outer asteroid belt the sheer proximity to Jupiter and the resulting strong short-term perturbations cause particles to be lost very rapidly.
Many of these regions may never have accumulated asteroids, and therefore the loss of these particles represents a numerical artifact in our simulation due to over-filling the model asteroid belt with test particles.
For instance, if the asteroid belt formed with the giant planets in their current positions, those regions would always have been unstable to asteroids, so none could have formed there.
However, regions of the asteroid belt that are currently highly unstable may not always have been.
Models of early solar system history indicate large changes to the orbital properties of the giant planets~\citep{Fernandez:1984p61,Hahn:1999p122,Tsiganis:2005p39}.
This ``numerical artifact'' is useful in indicating that the timescales of clearing in the strongly unstable zones is $\lesssim1\My$.
Below we discuss in detail the loss of asteroids from the more stable regions of the main belt.

\subsection{Historical population of large asteroids.}
We used the test particle loss history of Sim~1 to estimate the loss history of large asteroids from the main belt and the large asteroid impact rate on the terrestrial planets.
To do this we scaled $f$, the fraction of surviving particles in Sim~1 at $t=4\Gy$, to the number of large asteroids in the current main belt.
For this purpose, we define ``large asteroid'' as an asteroid with $D>30\km$. 
For most asteroids, size is not as well determined as absolute magnitude.
If the asteroid visual albedo, $\rho_v$, is known, the absolute magnitude can be converted into a diameter with the following formula~\citep{Fowler:1992p2510}:
\begin{equation}
D=\frac{1329\km}{\sqrt{\rho_v}}10^{-H/5}.
\label{e:rhotoD}
\end{equation}
Because asteroids can have a range of albedos, converting from brightness to diameter is fraught with uncertainty in the absence of albedo measurements. 
For simplicity, we adopt a single albedo, $\rho_v=0.09$, which is approximately representative over the size range of objects considered here~\citep{Bottke:2005p107}.
In subsequent analysis we will use absolute magnitude as a proxy for size.

Using Eq.~(\ref{e:rhotoD}) and our assumption of albedo, an asteroid of diameter $D=30\km$ has an absolute magnitude $H=10.8$.
The main belt is observationally complete for asteroid absolute magnitudes as faint as $H=13$~\citep{Jedicke:2002p17}.
While most $H\leq10.8$ asteroids have existed relatively unchanged over the last $4\Gy$, a few breakup events have created some large fragments over this timespan.
For example, there are five members of the Vesta family with $H<10.8$~\citep{Nesvorny:2006p2242}.
Collisional fragments produced over the last $4\Gy$ can ``contaminate'' the observed $H\leq10.8$ asteroid population, and collisional breakup events have also disrupted some primordial $H\leq10.8$ asteroids. 
These collisional processes complicate the estimate of the dynamical loss history of large asteroids over the age of the solar system.
Happily, most collisional family members with $H\leq10.8$ have been identified~\citep{Nesvorny:2006p2242}, and can be removed to further refine the observation sample.
The database of \citep{Nesvorny:2006p2242} was also made with some attempt to remove interlopers that have similar dynamical properties as a family, but a different spectral classification that indicates they are not members of the collisional family~\citep{MotheDiniz:2005p2512}

The observational data set we used was the 1137 asteroids with $H\leq10.8$ obtained from the AstDys online data service~\citep{Knezevic:2003p74}.
Using the family classification system of \cite{Nesvorny:2006p2242}, 206 ($18\%$) of these asteroids are identified members of collisional families.
We eliminated collisional family members and used the remaining sample of 931 asteroids for our scaling.
Implementing this normalization, Fig.~\ref{f:lhistory} shows the loss history of the main belt asteroids with the population scale on the right-hand axis.
Because the dynamical depletion of asteroids from the main belt is approximately logarithmic, a roughly equal amount of depletion occurred in the time interval $10$--$200\My$ as in $0.2$--$4\Gy$.
We find that the asteroid belt at $t=200\My$ would have had $28\%$ more large asteroids than today, and the asteroid belt at $t=10\My$ would have had $64\%$ more large asteroids than today.
Our calculation indicates that $\sim2200$ large asteroids ($H\leq10.8$) may have been lost from the main asteroid belt by dynamical erosion since the current dynamical structure was established, but $\sim1600$ of those asteroids would have been lost within the first $10\My$.

\subsection{Non-uniform pattern of depletion of asteroids}
\begin{figure}
\center
\resizebox{\textwidth}{!}{\includegraphics{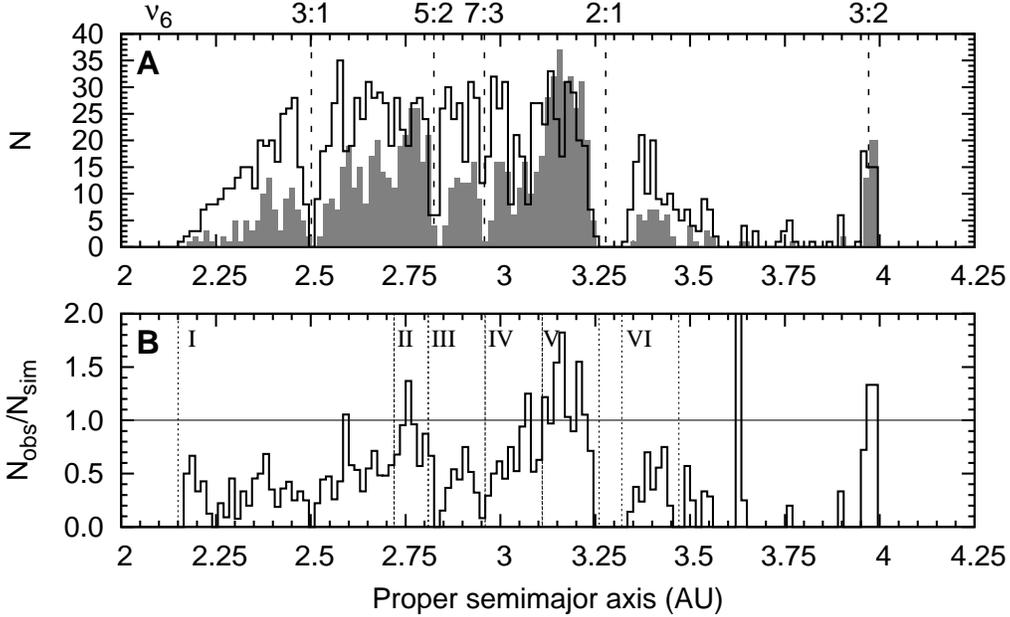}}
\caption{Comparison between the semimajor axis distribution of Sim~1 test particles and the sample of observed main belt asteroids.
\textbf{a)} The observed main belt asteroid distribution, $N_{obs}$, for $H\leq10.8$ asteroids (shaded) and the surviving particles of Sim~1, $N_{sim}$, (solid).
\textbf{b)} Ratio of the data sets.}
\label{f:depleted}   
\end{figure}

Fig.~\ref{f:depleted} compares the results of Sim~1 with the observational sample ($H\leq10.8$ asteroids, excluding collisional family members).\footnote{Fig.~\ref{f:depleted}a is similar to Fig.~1a of \cite{Minton:2009p280}, but with our sample of 931 asteroids with $H\leq10.8$ that are not members of collisional families. 
The results we report in this section are similar to those of \cite{Minton:2009p280}; because they are based on simulations with much larger number of particles, their statistical significance is improved.}
The proper semimajor axes of the surviving particles from Sim~1 at the end of the $4\Gy$ integration were computed using the public domain Orbit9\footnote{Found at: http://hamilton.dm.unipi.it/astdys/} code~\citep{Knezevic:2002p872}.
The bin size of $0.015\AU$ was chosen using the histogram bin size optimization method described by \cite{Shimazaki:2007p1189} (See~\ref{sec:asteroid_loss-appendix-optbinsize}).
Fig.~\ref{f:depleted}b is the ratio of the data sets. 
We find that the observed asteroid belt is overall more depleted than the dynamical erosion of an initially uniform population can account for, and there is a particular pattern in the excess depletion:
there is enhanced depletion just exterior to the major Kirkwood gaps associated with the 5:2, 7:3, and 2:1 mean motion resonances (MMRs) with Jupiter (the regions spanning $2.81$--$3.11\AU$ and $3.34$--$3.47\AU$ in Fig.~\ref{f:depleted}a); 
the regions just interior to the 5:2 and the 2:1 resonances do not show significant depletion (the regions spanning $2.72$--$2.81\AU$ and $3.11$--$3.23\AU$ in Fig.~\ref{f:depleted}a), but the inner belt region (spanning $2.21$--$2.72\AU$) shows excess depletion.

\cite{Minton:2009p280} showed that the observed pattern of excess depletion is consistent with the effects of the sweeping of resonances during the migration of the outer giant planets, most importantly the migration of Jupiter and Saturn. 
There is evidence in the outer solar system that the giant planets -- Jupiter, Saturn, Uranus and Neptune -- did not form where we find them today.  
The orbit of Pluto and other Kuiper Belt Objects (KBOs) that are trapped in mean motion resonances with Neptune can be explained by the outward migration of Neptune due to interactions with a more massive primordial planetesimal disk in the outer regions of the solar system~\citep{Malhotra:1993p244,Malhotra:1995p79}.  
The exchange of angular momentum between planetesimals and the four giant planets caused the orbital migration of the giant planets until the outer planetesimal disk was depleted of most of its mass, leaving the giant planets in their present orbits~\citep{Fernandez:1984p61,Hahn:1999p122,Tsiganis:2005p39}.  
As Jupiter and Saturn migrated, the locations of mean motion and secular resonances swept across the asteroid belt, exciting asteroids into terrestrial planet-crossing orbits, thereby greatly depleting the asteroid belt population and perhaps also causing a late heavy bombardment in the inner solar system~\citep{Liou:1997p75,Levison:2001p159,Gomes:2005p51,Strom:2005p80}.

\begin{figure}
\center
\resizebox{\textwidth}{!}{\includegraphics{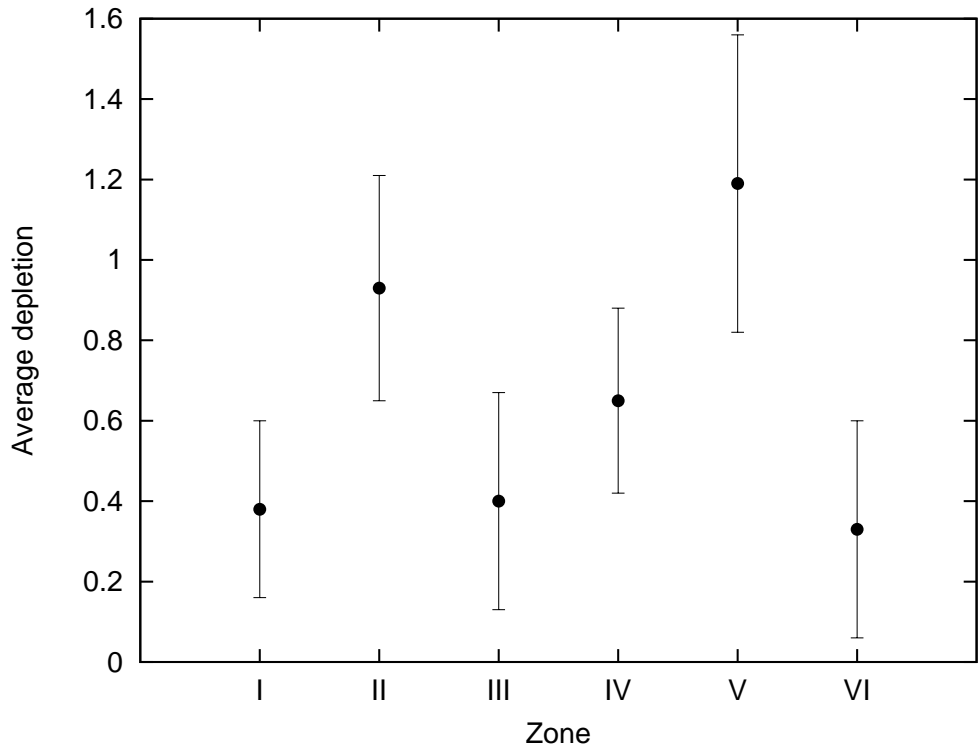}}
\caption{The average depletion, $\langle N_{obs}/N_{sim}\rangle$, in each of the six zones identified in Fig.~\ref{f:depleted}b; the average is taken over the $0.015\AU$ bins present in each zone. 
}
\label{f:zonedeplete}   
\end{figure}

We identified six zones of excess depletion; these are labeled I--IV in Fig.~\ref{f:depleted}b.
Zone I would have experienced depletion primarily due to the sweeping $\nu_6$ resonance, with some contribution possibly from the 3:1 MMR.
Zones II and IV are the zones that lie on the sunward sides of the 5:2 and 2:1 resonances, respectively, and are hypothesized to have experienced the least amount of depletion due to sweeping mean motion resonances (MMRs) and secular resonances under the interpretation of \cite{Minton:2009p280}.
Zones III, V, and VI are on the Jupiter-facing sides of the 5:2, 7:3, and 2:1 MMRs, respectively, and are hypothesized to have experienced depletion due to the sweeping of these resonances.
The average ratio between the model and observed population per $0.015\AU$ bin in each zone is quantified in Fig.~\ref{f:zonedeplete}.

\subsection{Empirical models of population decay}\label{subsec:empirical}
The loss rate of small bodies from various regions of the solar system has been studied by several authors~\citep[see][for a comprehensive review of the recent literature on the subject]{Dobrovolskis:2007p102}.
\cite{Holman:1993p3} found that the decay of a population of numerically integrated test particles on initially circular, coplanar orbits distributed throughout the outer solar system was asymptotically logarithmic, that is, $\dot{n}\propto t^{-1}$, where $n$ is the number of test particles remaining in the simulation at a given time $t$.  
\cite{Dobrovolskis:2007p102} showed that, for many small body populations, loss is described as a stretched exponential decay, given by the Kohlrausch formula,
\begin{equation}
	f=\exp\left(-\left[t/t_0\right]^\beta\right),
	\label{e:kohlrausch}
\end{equation}
where $f$ is the fraction remaining of the initial population ($f=n(t)/N_{tot}$).  
In the cases that \citeauthor{Dobrovolskis:2007p102} studied, namely loss rates of small body populations orbiting giant planets, they found that $\beta\approx0.3$.  
For reference we note that a value of $\beta=1/2$ is expected for a diffusion-dominated process for the removal of particles; in this case a plot of log of the number of remaining particles vs. square root time would be a straight line.

\begin{figure}[htb]
\center
\resizebox{\textwidth}{!}{\includegraphics{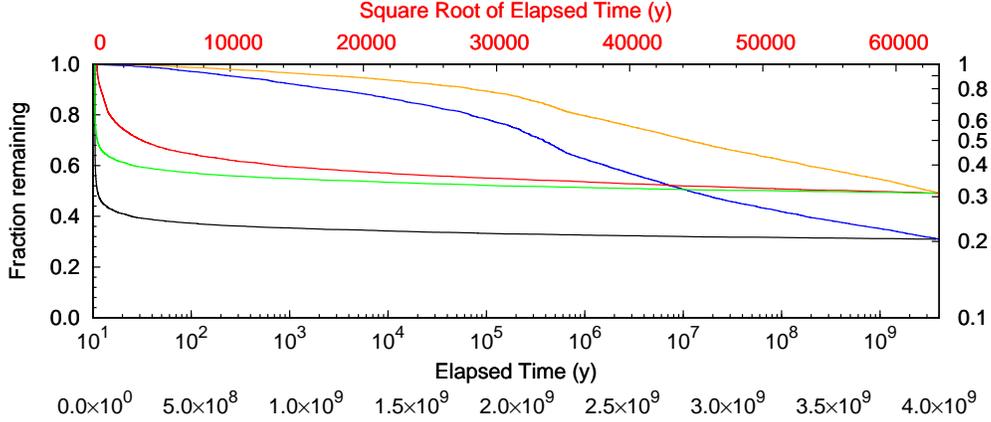}}
\caption{Comparison of empirical decay laws for the main asteroid belt region from Sim~1 (5760 particles for $4\Gy$). 
The result from Sim~1 is plotted in five different ways.  
Bottom curve (black): fraction $f$ of particles surviving (left-hand scale) vs. time (bottom scale).
Next-to-lowest curve (green): $\log f$ (right-hand scale) vs. elapsed time (bottom scale).
Middle curve (red): $\log f$ (right-hand scale) vs. $\sqrt{t}$ (top-scale).
Next-to-uppermost curve (blue): $f$ (left-hand scale) vs. $\log t$ (interior scale).
Top curve (yellow): $\log f$ (right-hand scale) vs. $\log t$ (interior scale).
Only the yellow and blue curves resemble straight lines in this format, and only for $t\gtrsim 10^6\yr$.}
\label{f:lhistory-dobrovolskis}      
\end{figure}

In Fig.~\ref{f:lhistory-dobrovolskis} we adopt a similar plot style as in \cite{Dobrovolskis:2007p102} (their Fig.~1) as a way of evaluating various empirical decay laws for the results of Sim~1.
Unlike the cases explored by \citeauthor{Dobrovolskis:2007p102}, stretched exponential decay with $\beta\sim 0.3$--$0.5$ is a very poor model for the asteroid belt.  
Fig.~\ref{f:lhistory-dobrovolskis} suggests either logarithmic or power law decay would be better models of particle decay from this simulation for $t>10^{6}\yr$.
Using a logarithmic decay law of the form:
\begin{equation}
	f=A-B\ln(t/1\yr),
	\label{e:logdecay}
\end{equation}
and a power law decay of the form:
\begin{equation}
	f=C (t/1\yr)^{-D},
	\label{e:powerdecay}
\end{equation}
where $A$, $B$, $C$, and $D$ are positive and dimensionless constants, we can look for the logarithmic, power law, and stretched exponential functions that best fit the decay history of Sim~1 for $t>1\My$.
The best fit parameters for each of these decay laws are listed in Table~\ref{t:decaylaws}.
Note that the best fit exponent $D$ for the power law decay is close enough to zero that it is not very different than the logarithmic decay over the range of timescales considered here.

\begin{table}[htb]
\begin{center}
\caption{Best fit decay laws for Sim~1 (5760 particles for $4\Gy$)}
\label{t:decaylaws}
\begin{tabular}{ l  l  l }
 \hline
Decay law & Parameters & Valid range (yr) \\
 \hline
 Stretched exponential (Eq.~\ref{e:kohlrausch}) & $\log t_0=8.6986\pm0.0057$ & $t>10^6$ \\
 & $\beta=0.1075\pm0.0004$ &  \\
 Logarithmic (Eq.~\ref{e:logdecay}) & $A=1.1230\pm0.0020$  & $t>10^6$ \\
 & $B=0.0377\pm0.0001$ & \\
 Power law (Eq.~\ref{e:powerdecay})& $C=1.9556\pm0.0027$ & $t>10^6$ \\
 & $D=0.0834\pm0.0001$ & \\
 \hline
 Piecewise logarithmic (Eq.~\ref{e:pwlogdecay})
 & $A_1=1.3333\pm0.0006$ & $10^{6.0}<t<10^{7.2}$\\
 & $B_1=0.05130\pm0.00004$ &  \\
 & $A_2=A_1+(B_2-B_1)\cdot7.2$ & $10^{7.2}<t<10^{8.3}$\\
 & $B_2=0.02695\pm0.00011$ & \\
 & $A_3=A_2+(B_3-B_2)\cdot8.3$ & $10^{8.3}<t<10^{9.1}$ \\
 & $B_3=0.02695\pm0.00011$ & \\
 & $A_4=A_3+(B_4-B_3)\cdot9.1$ & $10^{9.1}<t<10^{9.6}$ \\
 & $B_4=0.03079\pm0.00018$ & \\
 \hline
\end{tabular}
\end{center}
\end{table}

\begin{figure}[htb]
\center
\resizebox{\textwidth}{!}{\includegraphics{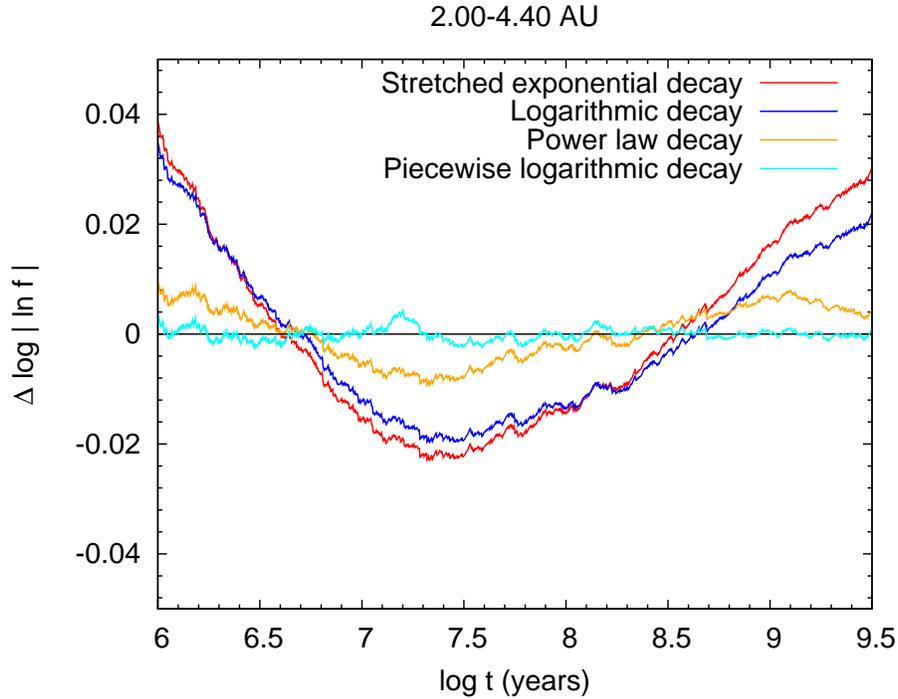}}
\caption{Differences between the best fit decay models and the loss history of Sim~1.
Here the y-axis is $\Delta\log|\ln f|=\log|\ln f_{sim}|-\log|\ln f_{fit}|$, where the subscripts $sim$ and $fit$ refer to the simulation data and best fit loss function, respectively. 
}
\label{f:kohlrausch}   
\end{figure}

The difference between various empirical decay laws and the simulation output from Sim~1 is shown in Fig.~\ref{f:kohlrausch}.
The format of Fig.~\ref{f:kohlrausch} is similar to that of Fig.~2 of \cite{Dobrovolskis:2007p102}, but here the y-axis is $\Delta\log|\ln f|=\log|\ln f_{sim}|-\log|\ln f_{fit}|$, where the subscripts $sim$ and $fit$ refer to the simulation data and best fit model, respectively.
A perfect fit would plot as a straight line with $\Delta \log|\ln f|=0$.
The power law function is a good fit, but with a small exponent $D$, which makes it practically similar to a logarithmic decay.  
The best fit stretched exponential value of $\beta$ obtained here is much smaller than the value of $\sim0.3$--$0.5$ obtained by many of the cases shown by \cite{Dobrovolskis:2007p102}.
This may indicate that classical diffusion does not dominate the loss of asteroids from the main belt.

From Fig.~\ref{f:kohlrausch} there is no clear preference for one or the other functions for the decay model.
However, with some experimentation, we found that an improved fit can be obtained by considering a piecewise logarithmic decay of the form:
\begin{equation}
	f_i=A_i-B_i\ln(t/1\yr),\; t_i<t<t_{i+1},
	\label{e:pwlogdecay}
\end{equation}
where $A_i$ and $B_i$ are positive coefficients.  
A physical justification for a piecewise logarithmic decay law is outlined in the following argument. 
If the region under study were divided into smaller subregions, and the loss of particles from each of those subregions follows a logarithmic decay law, then the linear combination of the decay laws for all subregions is the decay law for the total ensemble of particles, and is itself logarithmic.
However, if any subregion completely empties of particles, then that region remains empty (it cannot have negative particles), and so the decay law of that region no longer contributes to the decay law for the total ensemble of particles; the decay of the ensemble then undergoes an abrupt change in slope.
A piecewise logarithmic decay law for an ensemble of particles originating from the main asteroid belt region implies that the intrinsic loss rate from the asteroid belt is best described as $\dot{n}\propto t^{-1}$, but with different proportionality constants for different regions inside the belt.

We found that the model that minimized $\Delta\log|\ln f|$ for $t>10^6\yr$ is a four component piecewise logarithmic decay law with slope changes at $10^{7.4}\yr$, $10^{8.3}\yr$, and $10^{9.1}\yr$.
We performed a least-squares fit to the loss history of Sim~1, fitting it to the four component piecewise function given by Eq.~(\ref{e:pwlogdecay}); the best fit parameters are given in Table~\ref{t:decaylaws}.
The residuals for the piecewise logarithmic decay law are much reduced, compared to the other empirical models considered, as shown in Fig.~\ref{f:kohlrausch}.

\section{Large asteroid impacts on the terrestrial planets}
\defcitealias{BVSP:1981p1528}{BVSP,~1981}

\label{sec:asteroid_loss-impacts}
Although the impact history of the terrestrial planets is numerically dominated by small impactors, $D\lesssim10\km$,  the larger but infrequent impactors are also of great interest as they cause the more dramatic geological and environmental consequences. 
The dynamical origins of the latter have been less well studied because of the unavoidable small number statistics issues with them.  
Unlike the Yarkovsky effect, which is primarily responsible for populating the NEA population with $D\lesssim10\km$ asteroids, the dynamical chaos in the asteroid belt is a size independent process and dynamical erosion is the primary loss mechanism for large asteroids.  
\cite{Migliorini:1998p2277} investigated how $D>5\km$ asteroids from the main belt become terrestrial planet-crossing orbits, and found that weak resonance in the inner solar system were likely responsible for populating the NEOs.
We used our simulations to quantify the impact rates for the larger asteroidal impactors. 

As shown in \S\ref{subsec:empirical}, the long term dynamical loss of asteroids from the main belt is nearly logarithmic in time.  
The fate of any particular asteroid (the probability that it will impact a particular planet, the Sun, or be ejected from the solar system) is strongly dependent on its source region in the main belt~\citep{Morbidelli:1998p696,Bottke:2000p82}.  
For example, particles originating from the region near the $\nu_6$ secular resonance at the inner edge of the main belt have a $\sim1$--$3\%$ chance of impacting the Earth~\citep{Morbidelli:1998p696,Ito:2006p40}, whereas objects originating further out in the asteroid belt have much lower Earth impact probabilities~\citep{Gladman:1997p88}.
Our simulations indicate that large asteroidal impactors that enter the inner solar system may originate throughout the main belt, so we needed to compute the overall impact probabilities for impactors originating by dynamical chaos from the main belt as a whole. 
We do this by means of an additional numerical simulation that yields the terrestrial planet impact statistics for those particles of Sim~1 that were `lost' to the inner solar system. 
We then combine the impact probabilities with the $4\Gy$ loss history of large asteroids (Fig.~\ref{f:lhistory}) to estimate the number of large impacts onto the terrestrial planets.  
The details of these calculations are described below.  

\subsection{Impact probabilities\label{subsec:asteroid_loss-impacts-prob}}
\begin{figure}
\center
\resizebox{\textwidth}{!}{\includegraphics{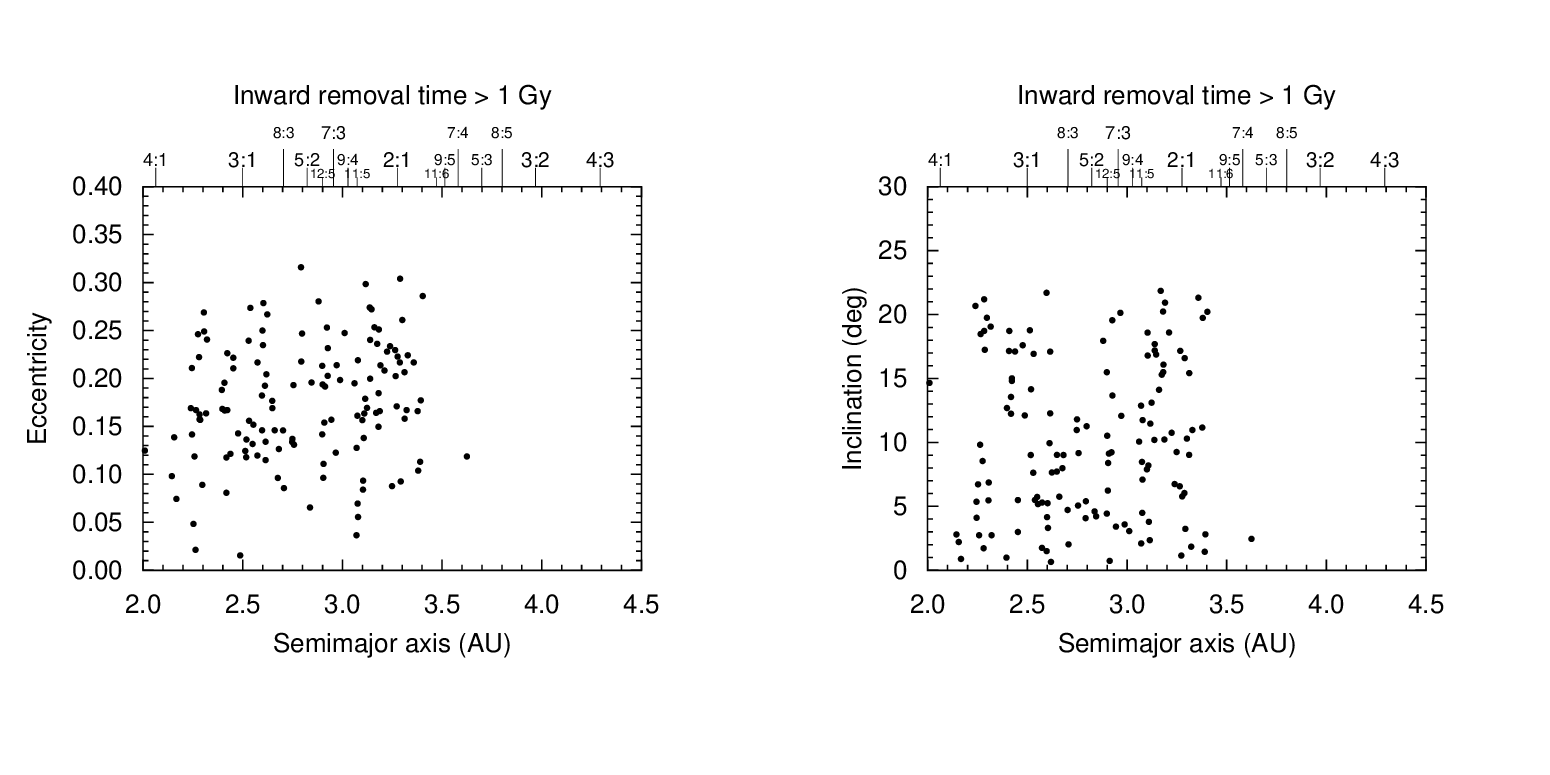}}
\caption{Distribution of initial orbital elements of the ``late'' particles from Sim~1.
These particles left the asteroid belt after $1\Gy$.
Only particles that were removed at an inward-going boundary are shown here, that is they were removed from Sim~1 either by crossing the inner barrier at $1\AU$ or crossing the Hill sphere of Mars.}
\label{f:cdist-inward}      
\end{figure}

In our initial simulations we did not follow any of the particles all the way to impact with any of the planets; particles that entered the inner solar system were stopped either at the Hill sphere of Mars or at an inner boundary of 1 AU heliocentric distance.  
To compute the impact probabilities for the terrestrial planets, we performed an additional simulation using the results of Sim~1.
Only inward-going particles from Sim~1 were considered, as outward-going ones are overwhelmingly likely to collide with or be ejected from the solar system by Jupiter, and as Figs.~\ref{f:cdist} and \ref{f:depleted}a illustrate, there are few observed asteroids beyond $3.4\AU$, where outward-going asteroids dominate.
We first identified a set of ``late'' particles from Sim~1 that were removed after $1\Gy$ at an inner boundary (either at the cutoff at $1\AU$ or by crossing the Hill sphere of Mars).
The Kirkwood gaps are mostly emptied of asteroids by $1\Gy$, as shown in Fig.~\ref{f:cdist}.
Fig.~\ref{f:cdist-inward} shows the initial semimajor axes, eccentricities, and inclinations of these 136 particles. 
Note from this figure that asteroids lost due to dynamical erosion after $1\Gy$ can come from nearly anywhere in the main belt.

First we captured the positions and velocities of the 136 late particles from Sim~1 at the time of their removal.
We cloned each of the late particles 128 times, such that $\mathbf{r}_{clone}=(1+\delta_r)\mathbf{r}_{original}$ and $\mathbf{v}_{clone}(1+\delta_v)\mathbf{v}_{original}$, where $|\delta_r|,|\delta_v|<0.001$.
The resulting 17408 particles were integrated using the MERCURY integrator with its hybrid symplectic algorithm capable of following particles through close encounters with planets~\citep{Chambers:1999p13}.
In this simulation, which we designate Sim~3, all eight major planets were included, the nominal integration step size was 2~days, and an accuracy parameter of $10^{-12}$ was chosen.
The particles were removed if they approached within the physical radius of the Sun or a planet, or if they passed beyond $100\AU$.
The simulation was run for $200\My$. 

For every particle that was removed in Sim~3 (either by impact or escape) we determined from which of the 136 source particles from Sim~1 it was cloned.
We used the initial semimajor axis of a given source particle and placed it into a bin of $0.015\AU$ in width, which we call the source bin. 
We then weighted the removal event by a factor equal to the ratio of the abundance of observed $H\leq10.8$ asteroids to the abundance of particles at the end of Sim~1 in the source bin.
The weighting factor, which also quantifies the relative amounts of depletion throughout the asteroid belt, is shown in Fig.~\ref{f:depleted}b.  
This weighting accounts for the differences in the orbital distribution of our simulated asteroid belt and the observed large asteroids. The raw impact statistics as well as the probabilities weighted based on the distribution of observed asteroids are tallied in Table~\ref{t:impactstats}.
These give the probability that an asteroid originating in the main belt, and which becomes a terrestrial planet-crosser by dynamical chaos, will impact a planet, the Sun, or be ejected from the solar system.
The unweighted and weighted terrestrial impact probabilities for the terrestrial planets differ by $<10\%$, and both give an impact probability onto the Earth of $\sim0.3\%$.

\begin{table}[htb]
\begin{center}
\caption{Impact probabilities for the terrestrial planets for Sim~3.  Ejected (OSS) refers to particles that either crossed the outer barrier at $100\AU$ or encountered the Hill sphere of a giant planet.}
\label{t:impactstats}
\begin{tabular}{ l c c c }
\hline
Fate & Number & \% &Weighted \%\\
\hline
Ejected (OSS)  & 13862 &  78.6 &  70.0\\
Survived &   269 &   1.55 &  13.7\\
Sun      &  3111 &  17.9 &  15.3\\
Mercury  &    10 &   0.057 &   0.061\\
Venus    &    74 &   0.425 &   0.396\\
Earth    &    56 &   0.322 &   0.306\\
Mars     &    26 &   0.149 &   0.140\\
\hline
\end{tabular}
\end{center}
\end{table}

\subsection{Flux of large (D$>$30~km) impactors on the terrestrial planets}
We used the weighted impact probabilities shown in Table~\ref{t:impactstats} and our model of the loss history of asteroids from Sim~1 to estimate the number of $D>30\km$ impacts onto the terrestrial planets since the end of the LHB.
Here we make the assumption that $t=0$ in our model is about $4\Gy$ ago, roughly the post-LHB era.
Because the loss rate of asteroids is approximately logarithmic, the loss rate is much higher at early times than later ones.
The early time is likely to have coincided with the tail end of the LHB itself; separating out the component of the impact flux that is due to the LHB rather than to dynamical erosion is problematic.  
We therefore consider only the dynamical loss at $t>100\My$ in our model as part of the post-LHB epoch.
With these assumptions, our model finds that since the end of the LHB, the Earth has experienced $\sim1$ impact of a $D>30\km$ asteroid. 
Venus, with its slightly higher impact probability than Earth, should have experienced $\sim1.3$ impacts of this size since the end of the LHB.
Our model also suggests that Mars and Mercury have had $\sim0.6$ and $\sim0.1$ impacts of $D>30\km$ asteroids, respectively.
These small numbers are consistent with there having been no impacts of $D>30\km$ asteroids on the terrestrial planets since the end of the LHB.

\subsection{Comparison with record of large impact craters on the terrestrial planets}
Known large impact basins on the terrestrial planets are generally of ages confined to the first $\sim800\My$ of solar system history, the Late Heavy Bombardment~\citep{Hartmann:1965p7,Ryder:2002p111}.
Excluding those, we consider only the post-LHB large craters on Earth and Venus.

The three largest known impact structures on Earth are Vredefort, Sudbury, and Chicxulub craters, each with final crater diameters $D_c<300\km$ and ages less than $\sim2\Gy$~\citep{Grieve:2008p204}.
\cite{Turtle:1998p141} argue that Vredefort crater has a diameter of $D_c<200\km$, which would make the final diameters of all of the largest known impact craters on Earth $D_c\sim130$--$200\km$.
At least one of these large impact events has been associated with a mass extinction.
The Chixculub crater, estimated to have been created by the impact of a $D\sim10\km$ object, is associated with the terminal Cretaceous mass extinction event~\citep{Alvarez:1980p1157,Hildebrand:1991p1162}.
Estimates from impact risk hazard assessments in the literature suggest that Chixculub-sized impact events happen on Earth on the order of once every $10^8\yr$~\citep{Chapman:1994p1290}.

The impact cratering record of Venus is unique in the solar system.
Over 98\% of the surface of Venus was mapped using synthetic aperture radar by the Magellan spacecraft~\citep{Tanaka:1997p977}.
The observed craters on Venus appear mostly pristine and are randomly distributed across the planet's surface, which has been taken as evidence for a short-lived, global resurfacing event on the planet within the last $\sim1\Gy$~\citep{Phillips:1992p1198,Strom:1994p1226}.
Venus has four craters with $D_c>150\km$; Mead crater with $D_c=270\km$, Isabella crater with $D_c=175\km$, Meitner with $D_c=149\km$, and Klenova with $D_c=141\km$.\footnote{From the USGS/University of Arizona Database of Venus Impact Craters at http://astrogeology.usgs.gov/Projects/VenusImpactCraters/}
\cite{Shoemaker:1991p781} estimated the surface age of Venus to be $\sim200$--$500\My$ using the total abundance of Venus craters and a flux of impactors based on the known abundance of Venus-crossing asteroids and on models of their impact probabilities.
More recently, \cite{Korycansky:2005p709} used similar techniques (as well as an atmospheric screening model for small impactors) and estimated the surface age of Venus to be $730\pm220\My$ old.
\cite{Phillips:1992p1198} used four methods to determine the age of Venus' surface, three which used models of observed Venus-crossing asteroids and one that used the observed abundance of craters on the lunar mare as a calibration.
All four methods resulted in surface ages between $400$-$800\My$.

Each of the techniques described above for estimating surface ages based on the abundance of observed craters has its shortcomings.  
Calculating a surface age using the observed population of NEAs makes the assumption that the current population of NEAs is typical for the entire post-LHB history of the inner solar system, including both the number of near earth asteroids and their computed impact probabilities.  
Calculating the surface age based on the abundance of craters on the lunar mare makes the assumption that the crater production rate in the inner solar system has been approximately constant over the last $\sim3.2\Gy$, which is the age by when most lunar mare were produced~\citepalias{BVSP:1981p1528}.

Large impactors produce craters with a complex morphology.  
To determine the sizes of impactors which created the largest post-LHB terrestrial impact craters, we used Pi scaling relationships to estimate the transient crater diameter as a function of projectile and target parameters; we then applied a second scaling relationship between transient crater diameter and final crater diameter.
The particular form of Pi scaling used here is given by \cite{Collins:2005p1232} for impacts into competent rock:
\begin{equation}
D_{tc}=1.161\left(\frac{\rho_i}{\rho_t}\right)^{1/3}D^{0.78}v_i^{0.44}g^{-.22}\sin^{1/3}\theta,
\label{e:piscale}
\end{equation}
where $\rho_i$ and $\rho_t$ are the densities of the impactor and target in kg$\m^{-3}$, $D$ is the impactor diameter in m, $v_i$ is the impactor velocity in m$\s^{-1}$, $g$ is the acceleration of gravity in m$\s^{-2}$ and $\theta$ is the impact angle.
We used the relationship between final crater diameter, $D_c$, and transient crater diameter, $D_{tc}$, given by \cite{McKinnon:1985p1258}:
\begin{equation}
D_c=1.17\frac{D_{tc}^{1.13}}{D_*^{0.13}},
\label{e:transient2final}
\end{equation}
where $D_*$ is the diameter at which the transition between simple and complex crater morphology occurs.
The transition diameter, $D_*$, is inversely proportional to the surface gravity of the target~\citep{Melosh:1989p151}, and can be computed based on the nominal value for the Moon of $D_{*,moon}=18\km$.

Applying Eqs.~(\ref{e:piscale}) and (\ref{e:transient2final}) to the problem of terrestrial planet cratering by asteroids requires some assumptions about both the impacting asteroids and the targets.
For simplicity we assumed that target surfaces have a density $\rho_t=3\gm\cm^{-3}$ and that impacts occur at the most probable impact angle of $45^\circ$~\citep{Gilbert:1893p1437}.
The characteristic impact velocity is often chosen to be the {\em rms} impact velocity obtained from a Monte Carlo simulation of planetary projectiles~\citepalias{BVSP:1981p1528}.
However, using the {\em rms} impact velocity may lead to misleadingly high estimates of the impact velocity because the impact velocity distributions are not Gaussian~\citep{Bottke:1994p2281,Ito:2009p1658}.
Therefore the median velocity may be more appropriate estimate of a ``typical'' impact velocity~\citep{Chyba:1990p89,Chyba:1991p1502}. 

\begin{figure}
\center
\resizebox{\textwidth}{!}{\includegraphics{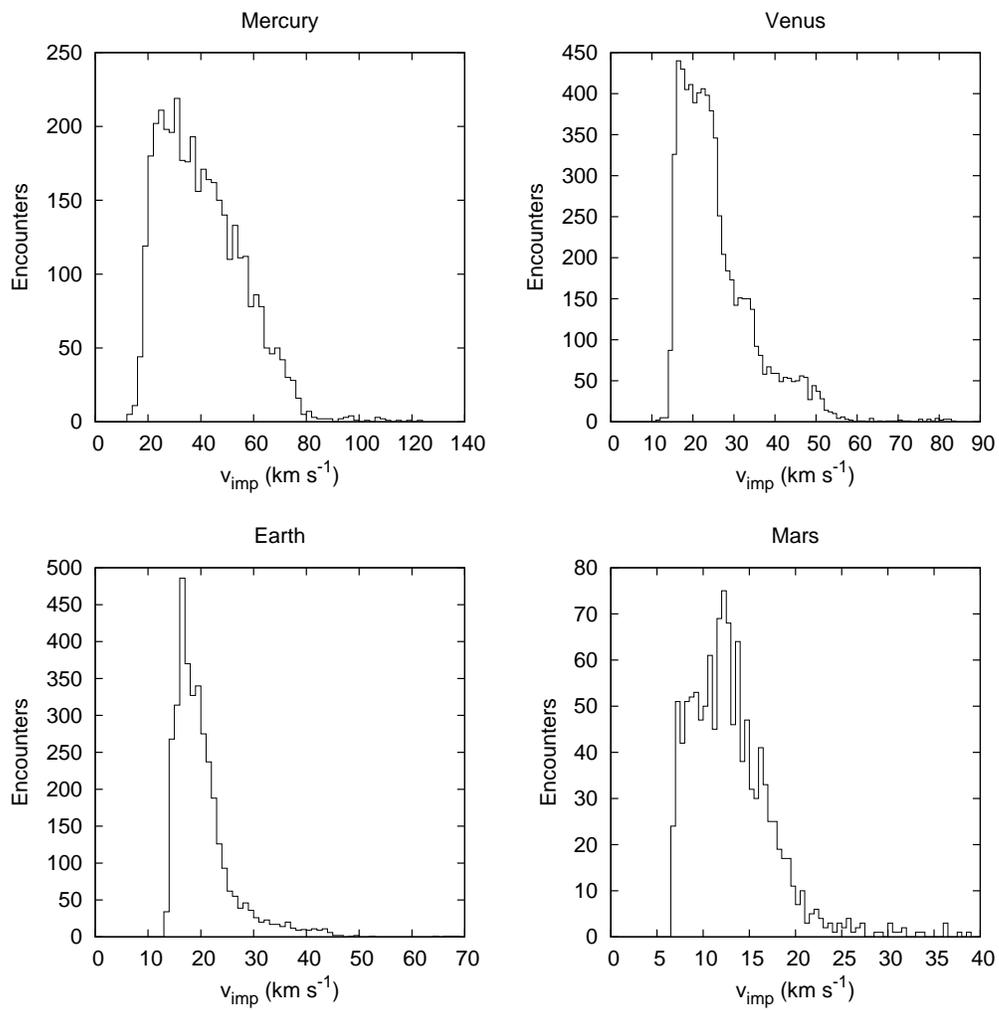}}
\caption{Impact velocity distributions of asteroids on the terrestrial planets.  }
\label{f:cehistory-impacts}      
\end{figure}

We made our own estimates of the impact velocities onto the planets using the results of Sim~3.
The small number of impacts computed in Sim~3 makes determining impact velocity distributions difficult.  
We improved the statistics by using the far more numerous close encounters between planets and test particles to calculate mutual encounter velocities.
For every close encounter in Sim~3 we recorded the closest-approach distance and mutual velocity between particles and planets. 
We only considered particles that had a closest-approach distance within the Hill sphere of a planet.
We estimated impact velocities using the {\em vis viva} integral given by:
\begin{equation}
\frac{1}{2}v_{imp}^2-\frac{GM_p}{r_p}=\frac{1}{2}v_{enc}^2-\frac{GM_p}{r_{enc}},
\label{e:visviva}
\end{equation}
where $v_{imp}$ and $r_p$ are the estimated impact velocity and the radius of the planet, and $v_{enc}$ and $r_{enc}$ are the mutual encounter velocity and closest approach distance.
Some particles encountered a planet multiple times, which skews the impact velocity estimates, so we only considered unique encounters between a particular particle and a planet.
If a particle encountered a planet multiple times it was treated as a single event and we only used the impact velocity of the first encounter.
The median, mean, and RMS impact velocities for each of the planets is shown in Table~\ref{t:velocities}.
The velocity distributions are shown in Fig.~\ref{f:cehistory-impacts}.

\begin{table}[htb]
\begin{center}
\caption{Estimated impact velocities for close encounter events in Sim~3.}
\label{t:velocities}
\begin{tabular}{ l c c c c }
\hline
Planet & Encounters & \multicolumn{3}{c}{Velocity (km$\s^{-1}$)} \\
 & & Median & Mean & RMS \\
\hline
Mercury &  3885 & 38.1 & 40.5 & 43.3\\
Venus &  6974 & 23.4 & 25.9 & 27.5\\
Earth &  3522 & 18.9 & 20.3 & 21.1\\
Mars &  1205 & 12.4 & 13.1 & 13.8\\
\hline
\end{tabular}
\end{center}
\end{table}

Using the median impact velocities and our assumptions about asteroid density ($\rho_i=1.5$--$3\gm\cm^{-3}$), we calculated the sizes of the impactors that produced the largest post-LHB craters in the inner solar system, using Eqs.~(\ref{e:piscale}) and (\ref{e:transient2final}).
The ``big three'' terrestrial impact craters are consistent with asteroidal impactors of diameter $D\sim7$--$14\km$. 
The estimated projectile size for Mead crater, the largest impact crater on Venus, is $D\sim12$--$16\km$.
Varying the assumptions used in the Pi scaling, Mead and Isabella craters are fully consistent with an impact of a $D>10\km$ asteroid, but Meitner and Klenova craters could be consistent with smaller asteroidal impacts.

In summary, there are no known impact structures of ages $\lesssim3.8\Gy$ (post-LHB) attributed to projectiles with $D\gtrsim30\km$.
This is consistent with the theoretical estimate above.
The small number of observed large impacts are consistent with the impact of $D>10\km$ objects.
Earth has been impacted by at least 3 objects that are consistent with $D>10\km$ asteroids.
During the Phanerozoic eon (the last $545\My$), only one impact crater, Chixculub, has been discovered on Earth that is consistent with a $D>10\km$ asteroid. 
Venus has been impacted by 2--4 $D>10\km$ objects in the last Gy, based on the number of observed large craters and its estimated surface age.

\subsection{Flux of D$>$10~km impactors on the terrestrial planets}
In our discussion of the dynamical erosion of the main asteroid belt, we have confined ourselves to $D>30\km$ primordial asteroids because non-gravitational and collisional effects are negligible for this population.
However, considering that the largest craters on the terrestrial planets correspond to impactors $D\sim10\km$, somewhat smaller than $30\km$, we are motivated to consider the $D\gtrsim10\km$ population of the main belt.
In the size range $D=10$--$30\km$ the effects of collisions and non-gravitational forces are not negligible, but they do not dominate that of dynamical chaos on the semimajor axis mobility of main belt asteroids.
Therefore we extend our dynamical calculations to $D>10\km$ asteroids and we compare the results with the terrestrial planet impact crater record, with the caveat that our results are only a rough estimate of the actual impact flux.

In order to turn the asteroid loss rate into an impact flux for a given crater size, the results of Sim~1, previously normalized to the abundance of $H\leq10.8$ asteroids (Fig.~\ref{f:lhistory}), were scaled to the fainter asteroids ($H<13.2$). 
Applying $\rho_v=0.09$ for the geometric albedo, $H<13.2$ corresponds to $D>10\km$.
The size distribution of the large asteroids of the main belt has not substantially changed over the last $\sim4\Gy$ and is described well by the present size distribution~\citep{Bottke:2005p107,Strom:2005p80}.
We took the $H\leq10.8$ ($D>30\km$) loss rate shown in Fig.~\ref{f:lhistory} and scaled it to $H<13.2$ ($D>10\km$) assuming the asteroid cumulative size distribution in this size range is a simple power law of the form:
\begin{equation}
	N_{>D}=kD^{-b}.
	\label{e:sfd}
\end{equation}
The debiased main belt asteroid size frequency distribution determined by \cite{Bottke:2005p107} gives $b=2.3$ in the size range $10\km<D<30\km$.

\begin{figure}
\center
\resizebox{\textwidth}{!}{\includegraphics{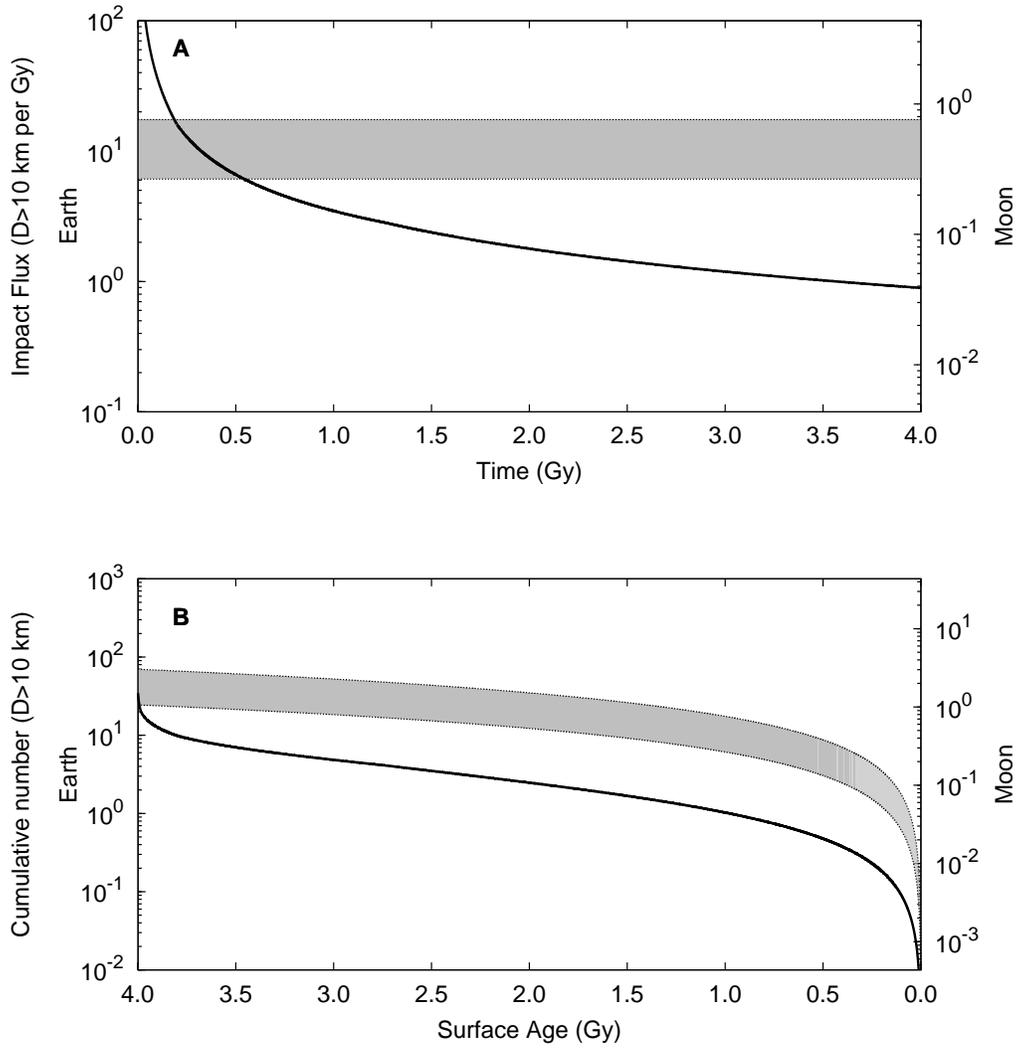}}
\caption{Impact rates of $D>10\km$ asteroids on the Earth and Moon.
{\bf a)} Solid line is the impact flux of $D>10\km$ impactors  per Gy onto the Earth and the Moon, from our calculations.
{\bf b)} Solid line is the cumulative number $D>10\km$ impactors for a given surface age, from our calculations.
The shaded regions indicate the Neukum Production Function~\citep{Neukum:2001p85} upper and lower bounds on the production of $D_c>140$--$200\km$ impact craters.
}
\label{f:num_basins}      
\end{figure}

We used our estimate of the loss of $D>10\km$ asteroids from the main belt due to dynamical diffusion, along with our estimate of the impact probabilities shown in Table~\ref{t:impactstats}, to determine the impact flux of $D>10\km$ asteroids, $\dot{N}_{>10\km}$, on the terrestrial planets.
The impact flux is based on the four-component piecewise decay law given by Eq.~(\ref{e:pwlogdecay}) with parameters given in Table~\ref{t:decaylaws}, which gives the fraction remaining as a function of time.
First the derivative of Eq.(\ref{e:pwlogdecay}) is taken, yielding $\dot{f}(t)$.
In order to convert from a fraction rate, $\dot{f}(t)$, to the flux of $D>10\km$ impacts, $\dot{N}_{>10\km}(t)$, we multiplied $\dot{f}(t)$ by the coefficient, $C_{>10\km}$, defined as:
\begin{equation}
	C_{>10\km}=\frac{931}{f(4\Gy)}\left(\frac{10}{30}\right)^{-2.3}p_i 
	\label{e:fractiontoflux}
\end{equation}
The first component of the coefficient is the constant, $931/f(4\Gy)$, which normalizes the fraction remaining such that the total number of asteroids remaining at $t=4\Gy$ is 931; the latter is the total number of $H<10.8$ ($D>30\km$) asteroids in the observational sample.
The next component is $(10/30)^{-2.3}$, which scales the results to $D>10\km$, as given by Eq.~(\ref{e:sfd}).
Finally, the coefficient was multiplied by the weighted probability $p_i$ that terrestrial planet-crossing asteroids impact a given planet.
For the Earth, the impact probability is $p_{Earth}=0.003$ as given in Table~\ref{t:impactstats}.
By multiplying $\dot{f}(t)$ by the coefficient $C_{>10\km}$ we obtained the flux of impacts by $D>10\km$ asteroids on the Earth as a function of time, $\dot{N}_{>10\km}$.
By further multiplying by $1/23$, using the ratio of 23:1 Earth:Moon impacts calculated by \cite{Ito:2006p40}, we also estimated the lunar flux.
The results are shown in Fig.~\ref{f:num_basins}a, showing our estimate of the impact flux of $D>10\km$ asteroids on the Earth (left-hand axis) and Moon (right-hand axis) since $1\My$ after the establishment of the current dynamical architecture of the main asteroid belt.

For comparison, we also plot in Fig.~\ref{f:num_basins}a the impact flux estimates obtained by~\cite{Neukum:2001p85} from crater counting statistics.
The shaded region of Fig.~\ref{f:num_basins}a represent upper and lower bounds on the estimated post-LHB impact flux using calibrated lunar cratering statistics~\citep{Neukum:2001p85}.
The lower bound is calculated using the number of $D_c>200\km$ lunar impact craters from the Neukum Production Function (NPF, Fig.~2 of \cite{Neukum:2001p85}); the rate is $7\times 10^{-9}$ craters$\km^{-2}\Gy^{-1}$. 
The upper bound is calculated using the number of $D_c>140\km$ lunar impact craters from the NPF; the rate is $2\times 10^{-8}$ craters$\km^{-2}\Gy^{-1}$.
These rates were multiplied by the surface area of the Moon to obtain the number of impacts on the lunar surface, and then multiplied by 23 to obtain the number of impacts on the Earth's surface.

We have also calculated the cumulative number of impacts on a surface with a given age; the result is shown in Fig.~\ref{f:num_basins}b.
The cumulative number of impacts as a function of time is calculated simply as $N_{cumulative}=C_{>10\km}\left[f(t)-f(4\Gy)\right]$.
Assuming that $t=0$ corresponds with an age of $4\Gy$ ago, the surface age is simply $SA=4\Gy-t$.
The upper and lower bounds on the cumulative number of impacts calculated from the NPF are shown as the shaded region, similar to Fig.~\ref{f:num_basins}a.
Our calculations of the impact flux show that at early times the flux was larger than the estimate given by the NPF, but that after $t=500\My$ the flux of impacts that we calculate is lower than that given by the NPF.
The present day flux of $D>10\km$ impactors is currently about an order of magnitude lower than that given by the NPF.

We note that the dynamical erosion rate decays with time, and the impact rate that we derive also decays as a function of time.
Our estimated impact flux for large craters declines by a factor of 3 over the last $3\Gy$, as shown in Fig.~\ref{f:num_basins}.
This is in some contrast with previous studies: in commonly used cratering chronologies, the impact flux is usually assumed to have been relatively constant over time since the end of the LHB.
The results obtained for the Earth-Moon system shown in Fig.~\ref{f:num_basins} can also be applied to the remaining terrestrial planets, using the weighted impact probabilities, $p_i$, given in Table~\ref{t:impactstats}.
Our estimated decay of the large impact rate for Mars over the last $4\Gy$ agrees with the results of \cite{Quantin:2007p119} that suggest that the impact cratering rate of $D_c>1\km$ craters on Mars has declined by a factor of 3 over the last $3\Gy$, based on counts of craters on 56 landslides along the walls of Valles Marineris.  
Reliable estimates are lacking for the absolute ages of lunar surfaces with ages $<3\Gy$, but what estimates do exist (i.e. estimated ages of Copernicus, Tycho, North Ray, and Cone craters from the Apollo missions) seem to be mostly consistent with a constant flux of impactors after $3\Gy$ ago, but with a possible increase in the flux after $1\Gy$ ago~\citep{Stoffler:2001p2}.
Also, estimates of the current impact flux from studies of the lunar record seem to be consistent with estimates made by observing the modern NEO population and estimating their impact probabilities~\citep{Morrison:2002p2438}.
However, \cite{Culler:2000p2407} suggested that the lunar impact flux declined by a factor of 2 or 3 from $\sim3.5\Gy$ ago until it reached a low at $\sim600\My$ ago and then increased again, based on dating of lunar impact glasses in soils.
This hypothesis is consistent with our result, that the overall impact flux onto the terrestrial planets has decreased by a factor of 3 since $\sim4\Gy$ ago due to the dynamical erosion of the asteroid belt.
The apparent increase in the flux at $\sim600\My$ ago until the present may be the result of a few large asteroid breakup events in the inner asteroid belt, such as the Flora and Baptistina family-forming events, and therefore the modern NEO population may not be representative of the NEO population over the last $\sim4\Gy$.

Assuming that $t=0$ corresponds to an age of $4\Gy$ ago, we estimate that in the last $3\Gy$ there have been $\sim4$ impacts of $D>10\km$ asteroids on the Earth due to dynamical erosion, and in the last $1\Gy$ there has been only 1 impact.
This is about an order of magnitude lower than estimates of the cratering rate of the terrestrial planets using the lunar cratering record, which give $\sim6$--$18$ $D>10\km$ impactors onto the Earth every $1\Gy$ using the NPF and assuming a $D>10\km$ object produces a $150$--$200\km$ crater~\citep{Neukum:2001p85,Chapman:1994p1290}.
The discrepancy between our estimates of the production of large craters and those based on the NPF may indicate that either (1) the current production rate of large craters ($D_c\gtrsim150\km$) is substantially overestimated by the NPF, or (2) the current production rate of large craters is not dominated by chaotic transport of large main belt asteroids.  

One possibility is that cratering at this size is dominated by comets, rather than asteroids.
The fraction of terrestrial planet impacts that are due to comets vs. asteroids has long been controversial, but is generally thought that Jupiter-family comets contribute fewer than $10\%$ and Oort cloud comets contribute fewer than $1\%$.~\citep{Bottke:2002p113,Stokes:2003p2316}.
Even periodic comet showers may not substantially increase the impact contribution from comets~\citep{Kaib:2009p1317}.
Therefore it seems unlikely that comet impacts can account for an order of magnitude more large impacts on the terrestrial planets than asteroids.

Another possibility is that large asteroid breakup events (followed by fragment transport via the Yarkovsky effect to resonances) dominate the production of large craters.
Our model neglects breakup events which have produced numerous $D\sim10\km$ fragments over the last $4\Gy$, and breakup events near resonances with Earth-impact probabilities higher than that of the intrinsic main belt may contribute to the large basin impact rate.
This is similar to the hypothesis proposed by \cite{Bottke:2007p719} for the origin of the Chicxulub impactor from the Baptistina family-forming event. 
The Baptistina breakup is hypothesized to have involved two large asteroids ($D_1\sim170\km$ and $D_2\sim60\km$) that collided at a semimajor axis distance $<0.01\AU$ from two overlapping weak resonances that, combined, increased fragment eccentricities to planet-crossing orbits with a $1.7\%$ Earth-impact probability~\citep{Bottke:2007p719}.
It is unclear whether such fortuitous combinations of conditions occur often enough to dominate the production of large craters over the solar system's history since the end of the LHB.
Large family forming breakup events do occur in the asteroid belt, and they likely can increase the flux of impacts onto the terrestrial planets~\citep{Nesvorny:2002p2310,Nesvorny:2006p2242,Korochantseva:2007p1706}.
The Flora family breakup event in particular was large and occurred in the inner asteroid belt roughly $0.5$--$1\Gy$ ago~\citep{Nesvorny:2002p2310}.
Also, while the Yarkovsky effect is very weak for $D>10\km$ asteroids, it is not entirely negligible over the age of the solar system.  
For instance, the loss rate of $D=10\km$ from the inner asteroid belt is somewhat higher when the Yarkovsky effect is taken into account compared to when it is not~\citep{Bottke:2002p117}.
It is doubtful whether this modest difference can account for a factor of ten increase in the flux of $D>10\km$ objects in the terrestrial planet region, however additional modeling may be needed to confirm this.
In addition, the weakness of the Yarkovsky effect on large asteroids can paradoxically enhance their mobility in the inner main belt.  
Combinations of nonlinear secular resonances and weak three-body resonances may cause asteroids to slowly diffuse through the middle and inner main belt~\citep{Carruba:2005p2513,Michtchenko:2009p2516}.
Only large asteroids that are only weakly affected by the Yarkovsky effect can remain inside these resonances long enough for them to act. 

\begin{figure}
\center
\resizebox{\textwidth}{!}{\includegraphics{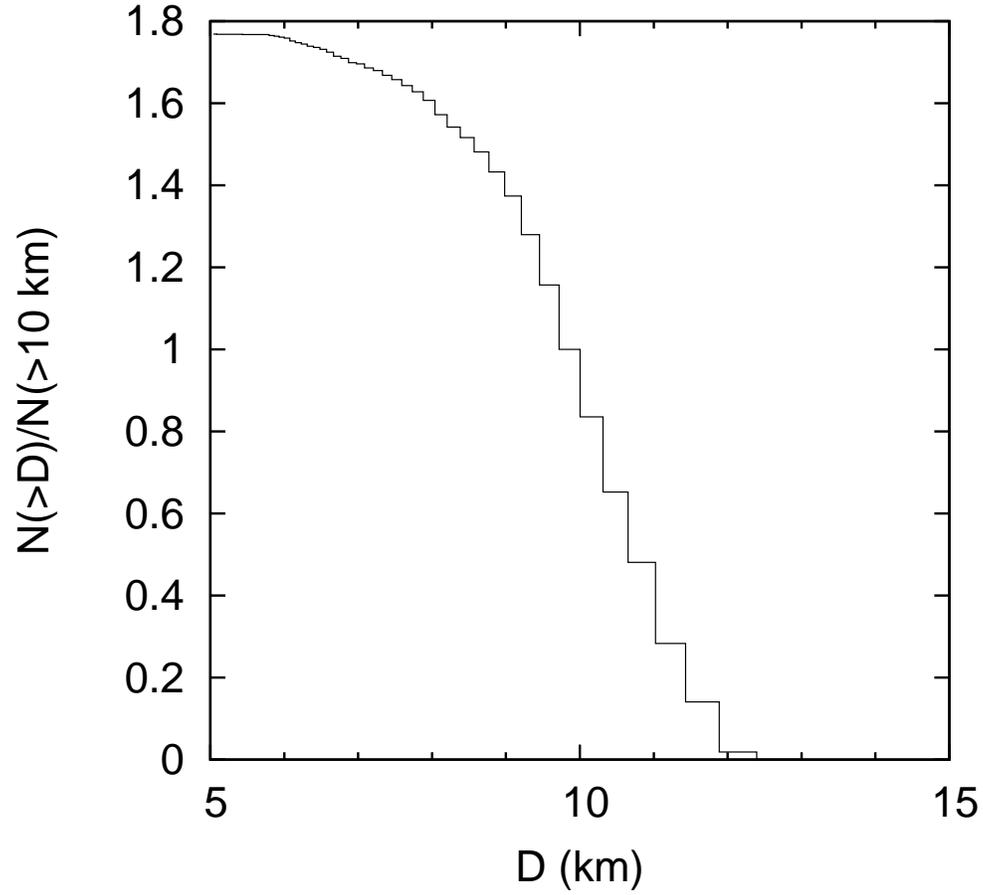}}
\caption{Cumulative size distribution of projectiles that contribute to the production of $D_c=187\km$ craters on Earth.
This is calculated by solving for the projectile size needed to produce a crater on Earth with a final diameter $D_c=187\km$ for the velocity distribution of Fig.~\ref{f:cehistory-impacts}c, and convolving the result with the asteroid size frequency distribution (Eq.~(\ref{e:sfd}), with index $b=2.3$).  This cumulative distribution is normalized such that the number of craters produced by objects of diameter $D>10\km$ is unity.
}
\label{f:vdist}      
\end{figure}

Finally, we consider the contribution from high velocity impactors with $D<10\km$ to the large impact crater production rate. 
The velocity distributions of asteroid impacts on the terrestrial planets have significant high velocity tails, as seen in Fig.~\ref{f:cehistory-impacts}.
Because size distributions of asteroids follow a power law with a negative index, as in Eq.~(\ref{e:sfd}), smaller objects are more numerous than larger ones.
We calculated the relative contribution of impacts by objects of varying sizes on the production of craters of a given size.
Using Eqs.~(\ref{e:piscale}) and (\ref{e:transient2final}), a $D=10\km$ projectile striking the Earth at $19\km\s^{-1}$ produces a crater with a final diameter $D_c=187\km$, assuming target and projectile densities are both $3\gm\cm^{-3}$, and the impact occurs at a $45^\circ$ angle.
We solved for the projectile size needed to produce a $D_c=187\km$ crater while varying the impact velocity.
The result was used to transform the Earth impact velocity distribution from Fig.~\ref{f:cehistory-impacts} into a projectile size distribution for a fixed final crater diameter.
Finally we multiplied the binned projectile size distribution by a binned asteroid size distribution, using a cumulative distribution as in Eq.~(\ref{e:sfd}) with an index $b=2.3$.
The result was then turned into a cumulative distribution and normalized to $N>10\km$, and is plotted in Fig.~\ref{f:vdist}.
This plot shows that asteroids with $D<10\km$ that impact at high velocity increase the production of large craters by no more than a factor of two. 
Therefore $D<10\km$ asteroids impacting at high velocity cannot account for the order of magnitude difference in the production rate of large impact craters on the Earth between our model and the NPF. 

On very ancient terrains, the discrepancy between total number of accumulated craters estimated from our model compared with the constant-flux models used in crater chronologies is less than with younger surfaces, as illustrated by Fig.~\ref{f:num_basins}b.
Our model also cannot account for the LHB itself.  
Total accumulated craters on ancient heavily cratered terrains associated with the LHB are at least $10$--$15$ times higher than on younger terrains~\citep{Stoffler:2001p2}, and likely even more if the surfaces reached equilibrium cratering.
The impact flux during the LHB was at least two orders of magnitude higher than the average flux over the last $3.5\Gy$, and possibly three orders of magnitude more if the LHB was a short-lived event~\citep{Ryder:2002p111}.
Even with the more enhanced rate of impacts at early times, and assuming higher Earth impact probabilities of $3\%$ based on estimates from the $\nu_6$ resonance, the impact flux due to dynamical erosion is roughly one or two orders of magnitude lower than that needed to produce the heavily cratered terrains associated with the LHB~\citep{Stoffler:2001p2,Neukum:2001p85}.

\section{Conclusion}
\label{sec:asteroid_loss-conclusion}
The main asteroid belt has unstable zones associated with strong orbital resonances with Jupiter and Saturn and relatively stable zones elsewhere. In most of the strongly unstable zones, the timescale for asteroid removal is $\lesssim 1$ Myr; one exception is the 2:1 mean motion resonance with Jupiter where the timescale to empty this resonance approaches a gigayear.  The relatively stable zones also lose asteroids by means of weak dynamical chaos on long timescales.  
We have found that the dynamical loss of test particles from the main asteroid belt as a whole is best described as a piecewise logarithmic decay. 
A piecewise logarithmic decay law implies that the intrinsic loss rate from the asteroid belt decays inversely proportional to time, $\dot{n}\propto t^{-1}$, but with different proportionality constants for different regions inside the belt.
When a region with a particular decay rate empties of asteroids, the decay law for the entire asteroid belt undergoes a change in slope.
This logarithmic loss of asteroids due to dynamical chaos was established very soon after the current dynamical architecture of the asteroid belt was established, and it continues to the present day with very little deviation.  Dynamical chaos is the predominant mechanism for the loss of large asteroids, $D\gtrsim10$--30 km.  
We have calculated that the asteroid belt $1\My$ after the establishment of the current dynamical architecture of the solar system had roughly twice its present number of large asteroids.
Because their loss rate is inversely proportional to time, we estimate that the flux of large impactors has declined by a factor of 3 over the last $3\Gy$.  We have calculated that large asteroidal impactors originating across the main asteroid belt have an overall Earth impact probability of $0.3\%$, and that the number of impacts of $D>10\km$ asteroids on Earth is only $\sim1$.
Our result on the current impact flux of $D>10\km$ asteroids due to dynamical erosion of the asteroid belt is an order of magnitude less than the values adopted in the recent literature on crater chronology and impact hazard assessment.
We have evaluated several possible explanations for the discrepancy and find them inadequate.
Our results can be used to improve studies of large impacts on the terrestrial planets.


\newcommand{\appsection}[1]{\let\oldthesection\thesection
  \renewcommand{\thesection}{Appendix \oldthesection}
  \section{#1}\let\thesection\oldthesection}
\begin{appendices}
\appsection{Determining the optimal histogram bin size}
\label{sec:asteroid_loss-appendix-optbinsize}
When presenting data as a histogram, the problem of what bin size to choose arises.
In the literature, the choice of bin size is often ad hoc, but need not be so.
In the case of the asteroid belt, we wish to analyze the semimajor axis distribution  of asteroids in order to gain insight into its past dynamical history.
If we choose a bin size that is too small, stochastic variation between bins can mask important underlying variations in the orbital element distributions.
If we choose a bin size that is too large, important small-scale variations become lost (i.e. the variability near narrow Kirkwood gaps).
In this work, we use histogram bin size optimizer developed by~\cite{Shimazaki:2007p1189} for optimizing time-series data with variability that obeys Poisson statistics.
This method is intuitive and easy to implement, and can be generalized to the problem of the number distribution of asteroids as a function of semimajor axis. 

\begin{figure}[htb]
\center
\resizebox{\textwidth}{!}{\includegraphics{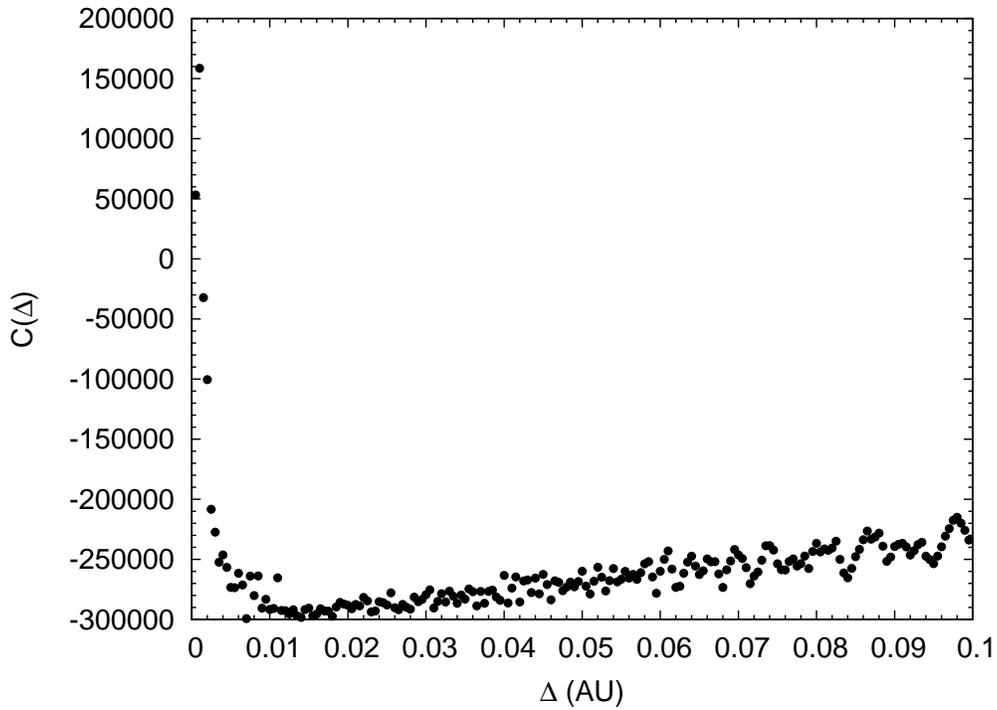}}
\caption{The cost function of Eq.~(\ref{e:binoptC}) as a function of the bin size $\Delta$ applied to the set of 931 asteroids with $H\leq10.8$ excluding collisional family members.
The minimum is nearly flat in the range of $0.01$--$0.02\AU$, so the choice of $0.015\AU$ is a reasonable one.
}
\label{f:Cvsbin-a}    
\label{lastfig}
\end{figure}

The optimal bin size is found by minimizing a cost function defined as:
\begin{equation}
C(\Delta)=\frac{2k-\nu}{\Delta^2},
\label{e:binoptC}
\end{equation} 
where the data have been divided into $N$ bins of size $\Delta$, $k$ is the mean of the number of asteroids per unit bin, and $\nu$ is the variance.
The mean is
\begin{equation}
k=\frac{1}{N}\sum_i^Nk_i,
\label{e:binoptk}
\end{equation}
and the variance is
\begin{equation}
\nu=\frac{1}{N}\sum_i^N(k_i-k)^2.
\label{e:binoptnu}
\end{equation}
The method works by assuming that the number of occurrences (or in our case, the number of asteroids), $k$, in each bin obeys a Poisson distribution such that the variance of $k$ is equal to the mean.
Fig.~\ref{f:Cvsbin-a} shows the $C(\Delta)$ for distribution in proper semimajor axis of our reduced set of $H\leq10.8$ asteroids described in \S\ref{sec:asteroid_loss-results}.
The optimal bin size is $\sim0.015\AU$ for this data set.
The results are similar for the surviving particles of Sim~1.

\end{appendices}

\section*{Acknowledgements}
The authors would like to thank H. Jay Melosh for his helpful comments regarding the impact flux calculation and crater size scaling.
The authors also thank Bill Bottke and Tatiana Michtchenko for their thoughtful and thorough reviews, and to David Nesvorn{\'y} for providing us with his catalogue of asteroid families.
This research was supported by NSF grant no.~AST-0806828 and NASA--NESSF
grant no.~NNX08AW25H.

\clearpage
\bibliographystyle{elsarticle-harv} 

\end{document}